%% file: main.tex
\documentclass[sigconf]{acmart}

\usepackage{multirow}
\usepackage{ragged2e}
\graphicspath{{./graphs/}}
\AtBeginDocument{%
  \providecommand\BibTeX{{%
    \normalfont B\kern-0.5em{\scshape i\kern-0.25em b}\kern-0.8em\TeX}}}

\copyrightyear{2022}
\acmYear{2022}
\setcopyright{acmcopyright}
\acmConference[ASHES'22]{Make sure to enter the correct
  conference title from your rights confirmation emai}{November 11, 2022}{Los Angeles, CA}
\acmDOI{}

\acmBooktitle{ACM Workshop on Attacks and Solutions in Hardware Security,
 November 11, 2022, Los Angeles, CA} 
\acmPrice{15.00}
\acmISBN{978-1-4503-XXXX-X/18/06}

\author{Muayad J. Aljafar}
 \affiliation{%Tallinn University of Technology
   \institution{Tallinn University of Technology}
   \city{ Tallinn}
   \country{Estonia}}
 \email{muayad.al-jafar@taltech.ee}
 
  \author{Florence Azais}
 \affiliation{%Tallinn University of Technology
   \institution{University of Montpellier}
   \city{ Montpellier}
   \country{France}}
 \email{florence.azais@lirmm.fr}
 
   \author{Marie-Lise Flottes}
 \affiliation{%Tallinn University of Technology
   \institution{University of Montpellier}
   \city{ Montpellier}
   \country{France}}
 \email{marie-lise.flottes@lirmm.fr}
 
   \author{Samuel Pagliarini}
 \affiliation{%Tallinn University of Technology
   \institution{Tallinn University of Technology}
   \city{ Tallinn}
   \country{Estonia}}
 \email{samuel.pagliarini@taltech.ee}

\begin{document}

\title{Leveraging Layout-based Effects for Locking Analog ICs}
%\email{webmaster@marysville-ohio.com}

%\affiliation{%
%  \institution{Institute for Clarity in Documentation}
%  \streetaddress{P.O. Box 1212}
%  \country{USA}

%\author{Lars Th{\o}rv{\"a}ld}
%\affiliation{%
%  \institution{The Th{\o}rv{\"a}ld Group}
%  \streetaddress{1 Th{\o}rv{\"a}ld Circle}
%\email{larst@affiliation.org}

%\affiliation{%
%  \institution{Inria Paris-Rocquencourt}

%\affiliation{%
% \institution{Rajiv Gandhi University}



\begin{abstract}
  While various obfuscation methods exist in the digital domain, techniques for protecting Intellectual Property (IP) in the analog domain are mostly overlooked. Understandably, analog components have a small footprint as most of the surface of an Integrated Circuit (IC) is digital. Yet, since they are challenging to design and tune, they constitute a valuable IP that ought to be protected. This paper is the first to show a method to secure analog IP by exploiting layout-based effects that are typically seen as undesirable detractors in IC design. Specifically, we make use of the effects of Length of Oxide Diffusion and Well Proximity Effect on transistor for tuning the devices’ critical parameters (e.g., $gm$ and $Vth$). Such parameters are hidden behind key inputs, akin to the logic locking approach for digital ICs. The proposed technique is applied for locking an Operational Transconductance Amplifier. In order to showcase the robustness of the achieved obfuscation, the case studied circuit is simulated for a large number of key sets, i.e., $>$50K and $>$300K, and the results show a wide range of degradation in open-loop gain (up to 130dB), phase margin (up to 50 deg), 3dB bandwidth ($\approx$2.5MHz), and power ($\approx$1mW) of the locked circuit when incorrect keys are applied. Our results show the benefit of the technique and the incurred overheads. We also justify the non-effectiveness of reverse engineering efforts for attacking the proposed approach. More importantly, our technique employs only regular transistors and requires neither changes to the IC fabrication process nor any foundry-level coordination or trust.
\end{abstract}



%%\ccsdesc[500]{Computer systems organization~Embedded systems}
%%\ccsdesc[300]{Computer systems organization~Redundancy}
%%\ccsdesc{Computer systems organization~Robotics}
%%\ccsdesc[100]{Networks~Network reliability}


\maketitle
\input{Introduction}
\input{Section2}
\input{Section3}

\input{Section4}

\input{conclusion}

\section{Acknowledgments}
This work has been partially conducted in the project “ICT programme” which was supported by the European Union through the European Social Fund. It was also partially supported by European Union’s Horizon 2020 research and innovation programme under grant agreement No 952252 (SAFEST).
\bibliographystyle{ACM-Reference-Format}
\bibliography{ashes}

\end{document}

%% file: Introduction.tex
\section{Introduction}\label{S1}

The semiconductor supply chain has been exposed to various security threats as a result of fabrication outsourcing. These threats are integrated circuit (IC) piracy, counterfeiting, overproduction, and hardware Trojans \cite{b1,b2,b3,b4}, causing a huge annual loss that was estimated a decade ago to be \$4 billion \cite{b5}. These security threats have been countered by design-for-trust  (DfTr) techniques, mostly applicable to digital ICs \cite{b6,b7}. Logic locking is a prime example of a DfTr technique \cite{b8}. 

However, the research effort for securing analog ICs or analog intellectual property (IP) is relatively small. Analog ICs are vulnerable to security threats as they have a small footprint and a wide range of use in nearly every application domain. Arguably, it is much easier to pirate analog ICs with a few hundred transistors than digital ICs with millions of transistors.
Previous works on analog logic locking aim at locking the circuits by key provisioning techniques\cite{b9} and tuning circuits functionalities\cite{b10} by either hiding the proper voltage or current bias, the transistor sizing, or the voltage thresholds of devices\cite{b11,b12,b13,b14,b15,b16,b17}. There are also other techniques used for locking analog mixed-signal (AMS) circuits where digital logic locking techniques have been applied for locking the digital part of the circuits \cite{b18,b19}. In addition, vulnerabilities of obfuscated analog circuits have been evaluated \cite{b20}, and attacks based on satisfiability modulo theories (\mbox{SMT}) and bias locking have been proposed \cite{b21,b22}. However, the technique we put forward in this paper relies on a completely different approach for analog obfuscation: \textbf{layout-based effects are leveraged to establish a key-based lock}. This is the first work to utilize this approach. 

\begin{figure}[b]
\centerline{\includegraphics[width=1.2\linewidth]{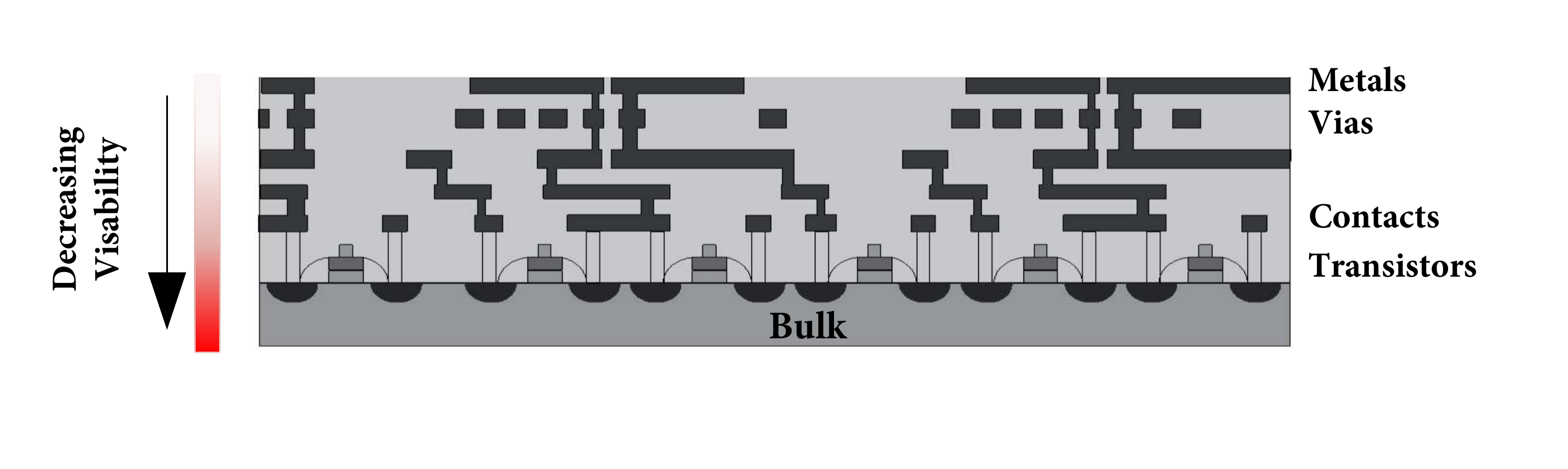}}
\caption{Cross section of the metal stack in an IC. Moving towards the base layer, the visibility decreases and the effort to reverse engineer increases.}
\label{fig: Fig. 1}
\end{figure}

We propose a technique for locking analog ICs that protects against counterfeiting and reverse-engineering (RE) attacks. Counterfeiting means illegally selling cloned ICs as original ones or selling illegitimately overproduced ICs in the aftermarket. RE techniques, on the other hand, are applied to derivate IC proprietary information such as its netlist and layout. The goal here is either to counterfeit the ICs (by extracting the information necessary for producing similar or identical ICs) or to steal secret information that an adversary should not be privy to. In RE techniques, the adversary needs to de-pack the IC, de-layer it, ``take pictures'' of the layers, and stitch the images together (likely by using specialized image processing tools) to obtain a netlist. While there are difficulties in this process (e.g., the number of individual images is barely tractable), the pictures clearly show the metal lines, vias, and even contacts. However, the features shrink as the delayering process gets closer and closer to the device’s layers that form the transistors. Finally, at the device level, doping gradients and other low-level properties are not trivial to obtain by delayering alone. A graphic notion of the difficulty in RE a complex metal stack is depicted in Fig. \ref{fig: Fig. 1}.

In this work, we perform obfuscation by manipulating two of such low-level properties in the diffusion layer; these are layout-based effects termed \textit{well proximity effect} (WPE) and \textit{length of diffusion} (LOD). This class of effects is often referred to as local layout effects or layout-dependent effects (LDEs), depending on the vendor. Surely, it is much harder to identify/characterize these effects than identifying the size of a transistor. To date, no RE attack has demonstrated this capability. In addition, extracting this level of detail is seemingly very expensive and time-consuming \cite{b23}. That being said, these effects have a direct impact on the transistors’ behavior such as the threshold voltage  ($Vth$) and transconductance ($gm$), which in turn affect the characteristics of an analog circuit. For example, for an operational transconductance amplifier (OTA), the effects would influence the power, gain, phase, and transconductance parameters.

%This paper capitalizes on these effects for locking analog circuits. Our simulations clearly show the impact of these layout-based effects on the power, gain, phase, and transconductance of an obfuscated .

The following contributions are made in this paper: 
\begin{itemize}
	\item It is shown, for the first time, how to capitalize on undesirable layout-based effects for locking analog circuits
	\item The effectiveness of the proposed technique is shown via a design example of an OTA
	\item We evaluate the effectiveness of our approach against RE efforts
\end{itemize}
The remainder of the paper is organized as follows. In Section II, the proposed technique is introduced and  explained. Section III shows a case study for the proposed locking technique and its results. Section IV discusses possible attack models and security analysis. Section V concludes the paper.

%% file: Section2.tex
\section{Background and Proposed Locking Technique}\label{S2}

\subsection{Layout-Dependent Effects}\label{S2.1}

\begin{figure}[t]
\centerline{\includegraphics [width=.95\linewidth]{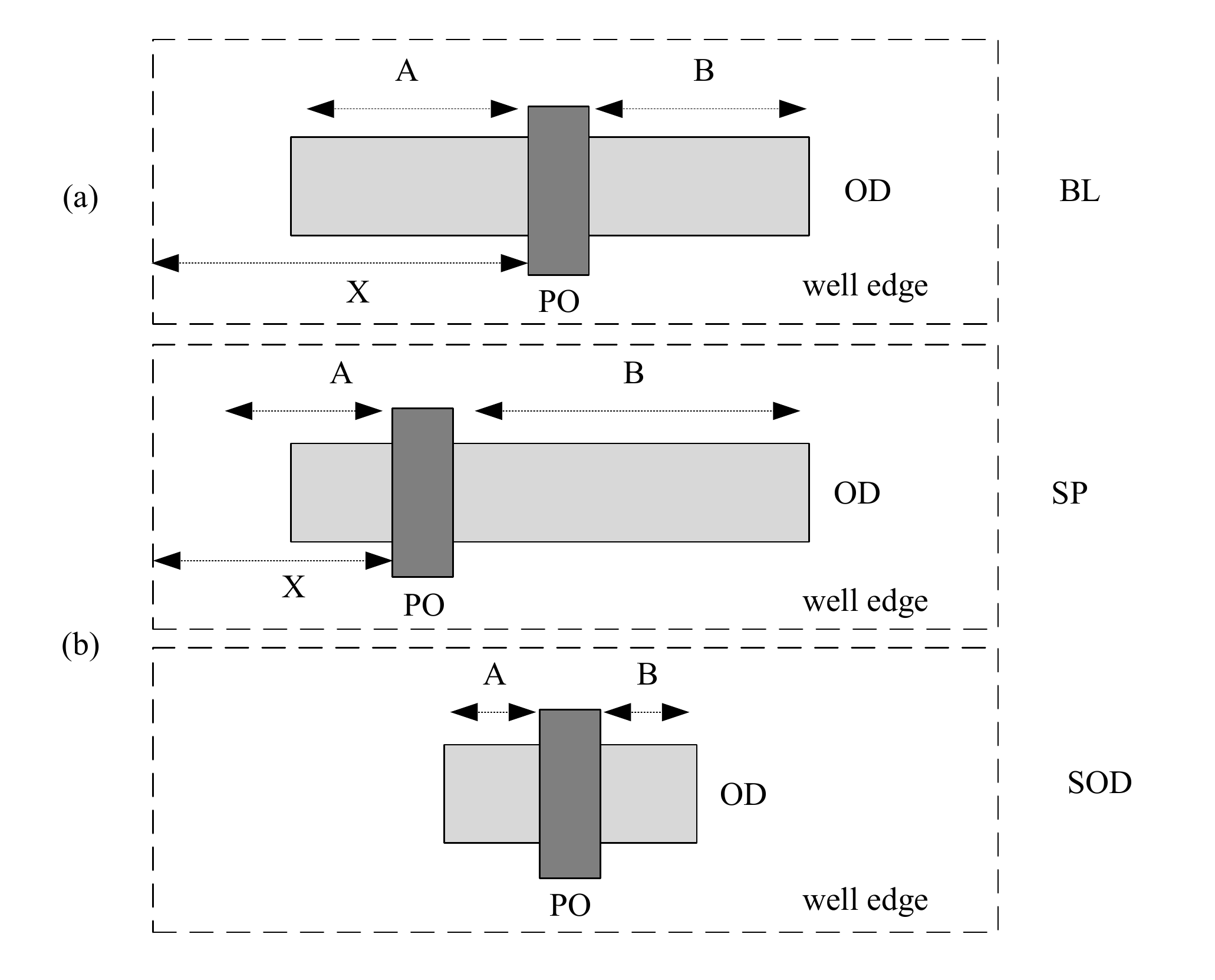}}
\caption{Layout-dependent effects. (a) Simplified transistor layout, baseline. (b) Different transistor arrangements, used for obfuscating analog circuits. OD is the oxide diffusion, and PO is the poly (gate). A and B are the distances between the poly and the OD edges, and X is the distance between the poly and the well edge. X relates to WPE, and A and B relate to LOD.}
\label{fig: Fig. 2}
\end{figure}

LDEs are a result of reducing the process geometries in lithography. WPE and LOD are examples of layout-dependent effects that appear in sub-100nm CMOS technology. However, we clarify that WPE and LOD effects are even more pronounced in technology nodes under 65nm. WPE relates to the device’s proximity to the well edge. A transistor that is close to the well edge will show a different performance (voltage threshold and drain current) from that of a device located far from the well edge (see X in Fig. \ref{fig: Fig. 2}a). This is due to the implant ions scattering off the resist side-well, even if the transistors are drawn with identical dimensions. LOD corresponds to different mechanical stress induced by a different OD length (i.e., poly to OD distances in Fig. \ref{fig: Fig. 2}a, marked as A and B), which affects the carrier mobility, hence the current in the devices. 

\begin{figure}[t]
\centerline{\includegraphics[width=.98\linewidth]{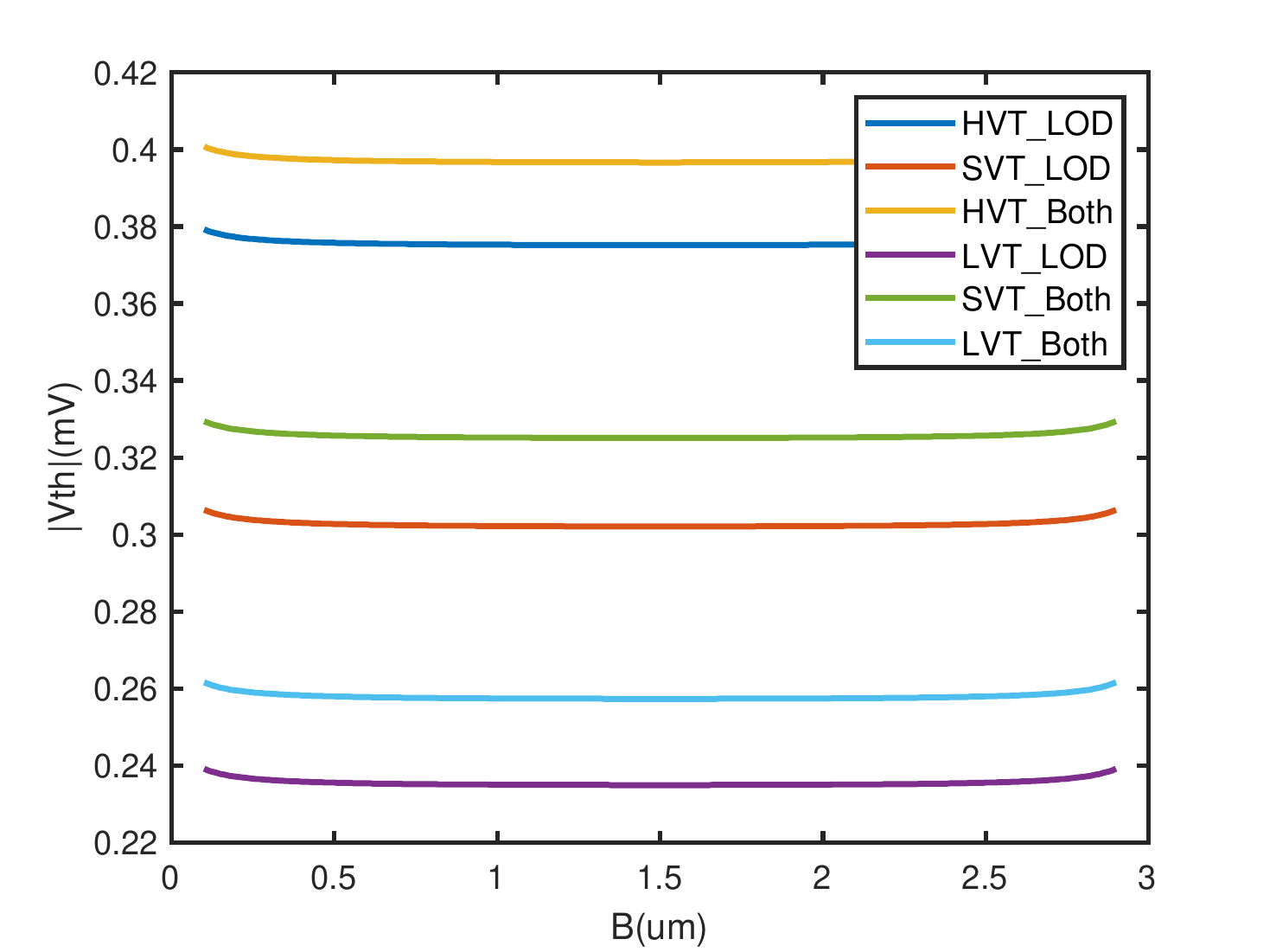}}
\centerline{\includegraphics[width=.98\linewidth]{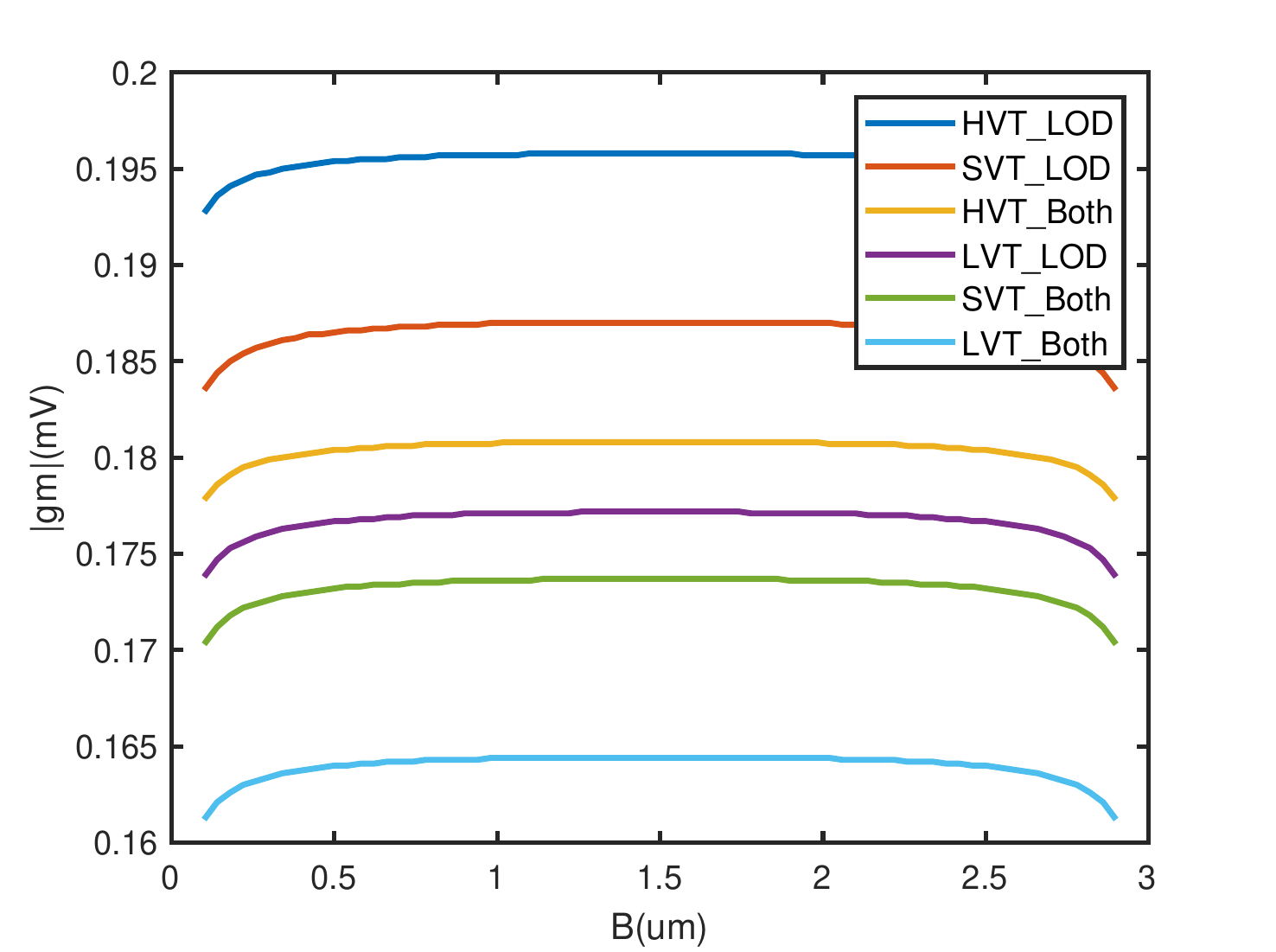}}
\caption{Effects of LOD and both LOD and WPE on the absolute values of voltage threshold and transconductance of PMOS transistors with a minimum length and representative width. B is shown in Fig. \ref{fig: Fig. 2}.}
\label{fig: Fig. 3}
\end{figure}

Fig. \ref{fig: Fig. 3} shows the effects of LOD and combined LOD and WPE on the absolute values of voltage threshold and transconductance of a PMOS transistor with standard (SVT-), high (HVT-), and low (LVT-) voltage thresholds at $v_{gs}=1V$.  For very small or very large B (Fig. \ref{fig: Fig. 3}), where the poly is close to the sides of the OD (Fig. {\ref{fig: Fig. 2}b}), the device shows different $V_{th}$ and $gm$ compared to other values for B. This will be exploited in this work for obfuscating analog ICs. These layout-dependent effects similarly impact the performance of an NMOS transistor. They also give rise to the device mismatch in analog circuits. 

In this work, our goal is to exploit these layout effects for locking analog circuits. To do so, we consider three arrangements for a transistor, namely baseline (BL), side-poly (SP), and short-OD (SOD), as shown in Fig. \ref{fig: Fig. 2}. BL corresponds to a nominal LDE case, whereas SP and SOD are utilized to further exploit WPE and LOD effects. With these arrangements, variations of up to $\approx$10\% in $Vth$ and $gm$ with respect to the BL can be obtained, as shown in Table ~\ref{tab:Table 1}. The voltage threshold variations in NMOS are larger than in PMOS, while transconductance variations in NMOS are smaller than in PMOS. The statistical variations (due to both process and mismatch) for all arrangements were also simulated. Table ~\ref{tab:Table 2} shows the standard deviations (SD) of $Vth$ and $gm$ with respect to their mean values. The numbers reported in Table ~\ref{tab:Table 1} and Table ~\ref{tab:Table 2} demonstrate that the layout-based effects are deterministic, i.e., they still present themselves no matter where the fabricated IC fares in the process variation spectrum.

%%%%%%%%%%%%%%%%%%%%%%%%%%%%%%%%%%%%%%%%%%%Fig. 4 %%%%%%%%%%%%%%%%%%%%%%%%%%%%%%%%%%%%%%%%%%
\begin{figure}[t]
%\vspace{-4mm}
\centerline{\includegraphics[width=.98\linewidth]{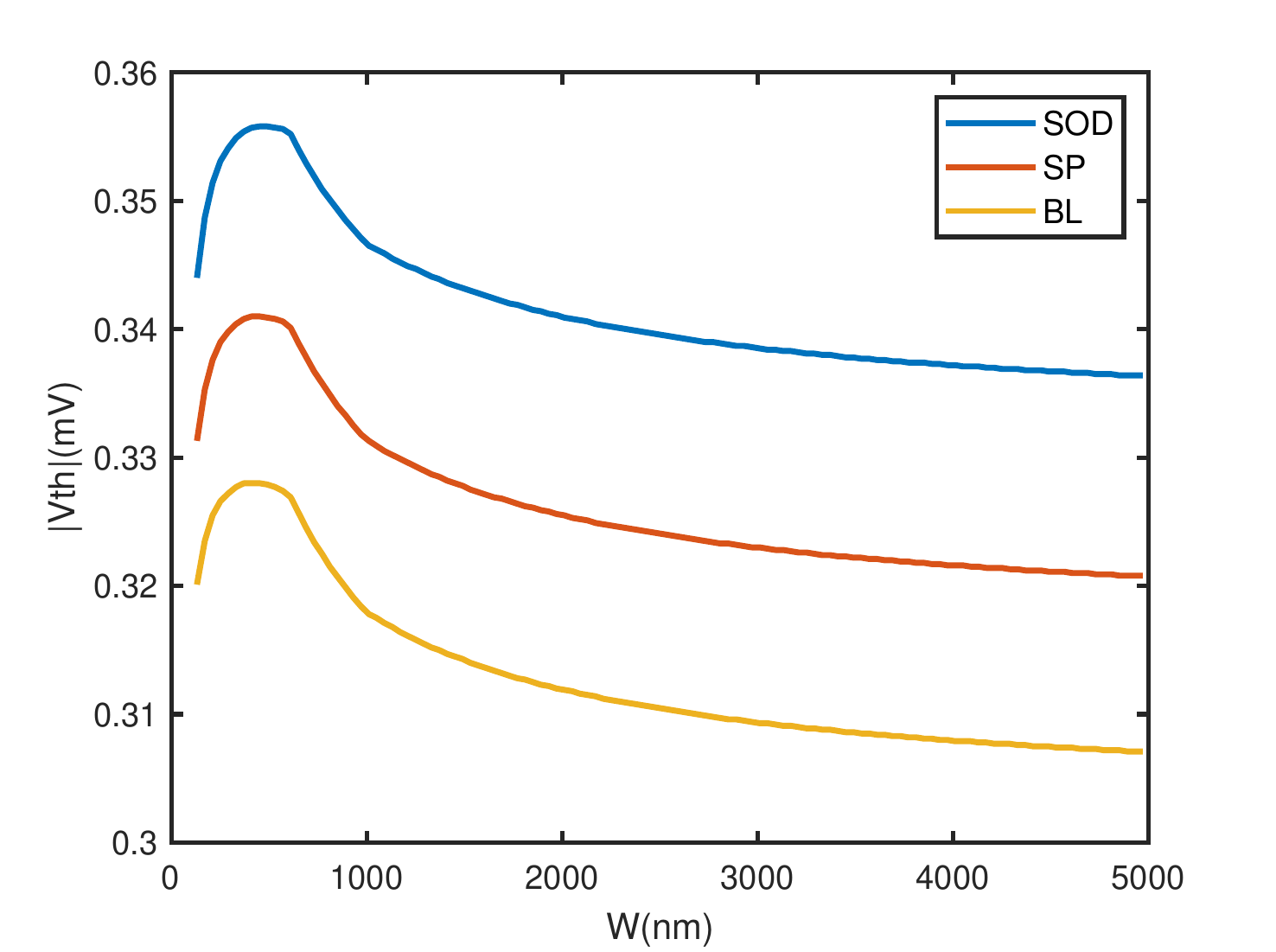}}
%\vspace{-1mm}
\caption{Effects of PMOS width, WPE and
LOD on the absolute values of voltage thresholds for all arrangements.}
\label{fig: Fig. 4}
\end{figure}

The effect of transistor width $W$ along with the layout-dependent effects on the voltage thresholds for all arrangements is shown in Fig. \ref{fig: Fig. 4} where the PMOS transistors have minimum length. Note that the margin between the lines BL-SP and BL-SOD is nearly constant, indicating that transistors of any size are potential candidates for obfuscation. In this example, the increase of W in SOD can change the voltage threshold variations from $6.8\%$ to $8.7\%$ compared to BL.
%%%%%%%%%%%%%%%%%%%%%%%%%%%%%%%%%%%%%%%%%%%%%%%% Table 1%%%%%%%%%%%%%%%%%%%%%%%%%%%%%%%%%%%%%%%%%%%%%%%%%%%%%%%% 
\begin{table}[t]
%\vspace{1mm}
	\caption{Variations (\%) in Voltage Threshold and Transconductance with respect to BL (the baseline)}
 	%\vspace*{-2mm}
	\begin{center}
		\begin{tabular}{c c c c c c c}
			\hline
			\hline
            \multirow{2}{*}{} &         					 & \multirow{2}{*}{$A_i$}   & \multirow{2}{*}{\sc Device}  &  \multicolumn{3}{c}{\sc Variations}    \\
            \cline{5-7}
	                          &          					 &        				    &                              & {\sc hvt(\%)} &{\sc svt(\%)} & {\sc lvt(\%)}  \\
	        \hline
            \parbox[t]{2mm}{\multirow{8}{*}{\rotatebox[origin=c]{90}{\sc parameter}}}       
                              & \parbox[t]{2mm}{\multirow{4}{*}{\rotatebox[origin=c]{90}{\sc $V_{th}$}}} 
                                                            & \multirow{2}{*}{\sc SP}   & {\sc pmos}                   & {\sc 2.85}    & {\sc 3.7}     & {\sc 4.59}     \\
            \cline{4-7}
			                 &          			        &         					& {\sc nmos}                   & {\sc 4.05}   & {\sc 4.38}    & {\sc 5}        \\
            \cline{4-7}
			                 &        			            & \multirow{2}{*}{\sc SOD} 	& {\sc pmos}                   & {\sc 6.08}   & {\sc 7.9}     & {\sc 9.79}     \\
	        \cline{4-7}
			                 &          				    &        				    & {\sc nmos}                   & {\sc 8.53}   & {\sc 9.28}    & {\sc 10.61}    \\
	        \cline{2-7}
                             & \parbox[t]{2mm}{\multirow{4}{*}{\rotatebox[origin=c]{90}{\sc $gm$}}}  	 
                                                            & \multirow{2}{*}{\sc SP}  & {\sc pmos}                   & {\sc 4.76}    & {\sc 4.72 }   & {\sc 4.68}    \\
	        \cline{4-7}
			                 &          				    &         					& {\sc nmos}                   & {\sc 1.72}    & {\sc 2.54}    & {\sc 2.42}    \\
	        \cline{4-7}
		                     &            		            & \multirow{2}{*}{\sc SOD} 	& {\sc pmos}                   & {\sc 10.4}    & {\sc 10.19 }  & {\sc 10.16}   \\
            \cline{4-7}
	     	                 &           		            &        					& {\sc nmos}                   & {\sc 3.7}     & {\sc 5.41}    & {\sc 5.09}     \\
			\hline
			\hline						
		\end{tabular}
	\label{tab:Table 1}
	\end{center}
	\small\justifying{Values were obtained from corner analysis for typical corner for devices with a minimum length and representative width. $A_i$ is an arrangement as defined in Fig. \ref{fig: Fig. 2}}
%	\vspace{-6mm}
\end{table}

%%%%%%%%%%%%%%%%%%%%%%%%%%%%%%%%%%%%%%%%%%%%%%%  Table 2  %%%%%%%%%%%%%%%%%%%%%%%%%%%%%%%%%%%%%%%%% 
\begin{table}[t]
%\vspace*{2mm}
	\caption{Process and mismatch of the arrangements}
	\begin{center}
		\begin{tabular}{c c c c c c c}
			\hline
			\hline
			\multirow{2}{*}{} &          & \multirow{2}{*}{$A_i$}   & \multirow{2}{*}{\sc Device}  &  \multicolumn{3}{c}{\sc Variations in SD*}     \\
			\cline{5-7}
		                      &      	 &        				    &                              & {\sc hvt(\%)} & {\sc svt(\%)} & {\sc lvt(\%)}  \\
			\hline
			\parbox[t]{2mm}{\multirow{12}{*}{\rotatebox[origin=c]{90}{\sc parameter}}}       
			                  & \parbox[t]{2mm}{\multirow{6}{*}{\rotatebox[origin=c]{90}{\sc $V_{th}$}}}   
			                             & \multirow{2}{*}{\sc BL} 	& {\sc pmos}                   & {\sc 9.78}     & {\sc 10.64}   & {\sc 12.95}   \\
			\cline{4-7}
			                  &          &         					& {\sc nmos}                   & {\sc 15.34}    & {\sc 12.29}   & {\sc 9.73}    \\
			\cline{4-7}
			                  &        	 & \multirow{2}{*}{\sc SP} 	& {\sc pmos}                   & {\sc 9.38}     & {\sc 10.12}   & {\sc 12.17}   \\
			\cline{4-7}
		                      &          &        				    & {\sc nmos}                   & {\sc 14.07}    & {\sc 11.28}   & {\sc 9.16}    \\
			\cline{4-7}
		                      &     	 & \multirow{2}{*}{\sc SOD}	& {\sc pmos}                   & {\sc 8.97}     & {\sc 9.58}    & {\sc 11.37}   \\
			\cline{4-7}
		                      &      	 &        				    & {\sc nmos}                   & {\sc 12.88}    & {\sc 10.34}   & {\sc 8.61}    \\
			\cline{2-7}
		                      & \parbox[t]{2mm}{\multirow{6}{*}{\rotatebox[origin=c]{90}{\sc $gm$}}}	 
			                             & \multirow{2}{*}{\sc BL}	& {\sc pmos}                   & {\sc 3.55}     & {\sc 3.98 }   & {\sc 2.90}    \\
			\cline{4-7}
		                   	  &      	 &         					& {\sc nmos}                   & {\sc 3.78}     & {\sc 2.85}    & {\sc 5.93}    \\
			\cline{4-7}
			                  &          & \multirow{2}{*}{\sc SP}	& {\sc pmos}                   & {\sc 3.35}     & {\sc 3.94 }   & {\sc 2.83}   \\
			\cline{4-7}
			                  &    		 &        					& {\sc nmos}                   & {\sc 3.63}     & {\sc 2.84}    & {\sc 5.98}     \\
			\cline{4-7}
		                      &    		 & \multirow{2}{*}{\sc SOD}	& {\sc pmos}                   & {\sc 3.17}     & {\sc 3.91 }   & {\sc 2.75}   \\
			\cline{4-7}
		                      &          &        					& {\sc nmos}                   & {\sc 3.45}     & {\sc 2.85}    & {\sc 6.05}     \\
			\hline
			\hline
			
		\end{tabular}	
		\vspace{2pt}
		\item \small {\centering *SD means standard deviation.}
		\label{tab:Table 2} 
	\end{center}
\end{table}
%%%%%%%%%%%%%%%%%%%%%%%%%%%%%%%%%%%%%%%%%%%%%%%%%%%%%%%%%%%%%%%%%%%%%%%%%%%%%%%%%%%%%%%%%%%%%%%%%%%%%%%%%%%%%

%In this work, we make use of three arrangements for obfuscating analog circuits, as shown in Fig. \ref{fig: Fig. 2}. However, distances X, A, and B are not discrete variables, i.e., there are many valid values for them that can be utilized to further obfuscate a design. In technology nodes under 65nm, many other LDEs exist -- those can also be harvested for obfuscation purposes.

\begin{figure}[t]
%\vspace{-3mm}
\centerline{\includegraphics[width=.75\linewidth]{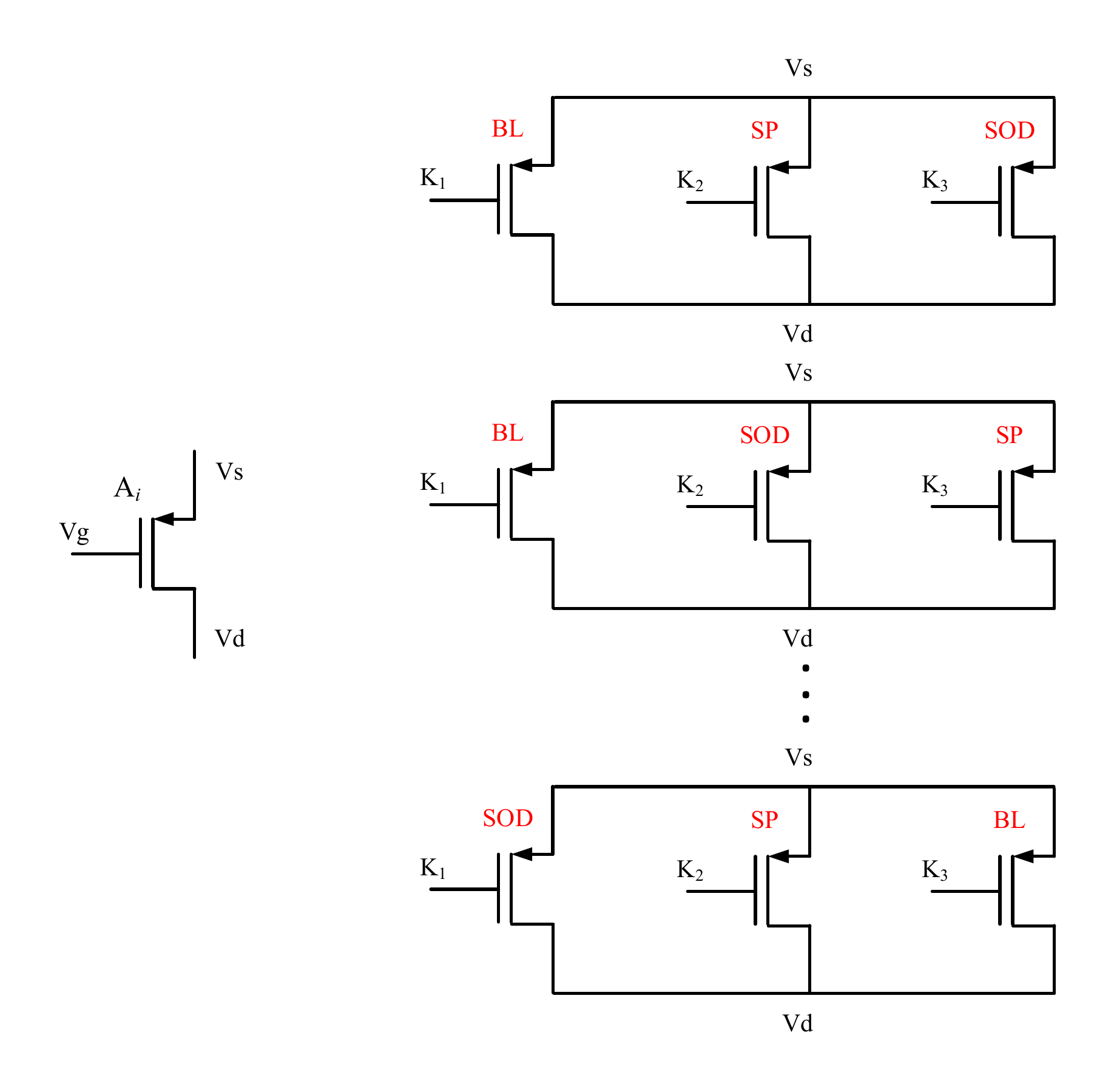}}
%\vspace{-5mm}
\caption{Principle of locking analog circuit. The order of these arrangements in the layout design is arbitrary. The figure shows only 3 out of 6 possible orders.}
\label{fig: Fig. 5}
\end{figure}

We propose designing analog circuits with different arrangements of transistors from which the `correct' arrangement is selected via keys. In principle, there are three possible arrangements for each NMOS or PMOS transistor where the order of these arrangements in a layout is arbitrary (e.g., SOD-BL-SP or SP-BL-SOD), where different orders lead to different correct key values. The three arrangements are controlled by three key bits (see Fig. \ref{fig: Fig. 5}). This procedure is the same for NMOS and PMOS transistors. Therefore, for a circuit with N devices, the keylength for the entire circuit is $N\times3$. In principle, there are a total of $2^{3N}$ possible combinations of arrangements or keys when assuming the keys are binary signals. 

However, it is observed that some of the `wrong' combinations lead to desirable performance, while others may produce nearly correct or completely incorrect performance/behavior. To explore this space and efficiently obfuscate an analog IP, we first propose a very simple three-step procedure as follows:

\begin{enumerate}
\item Design a circuit with a combination of BL/SP/SOD transistors
\item Examine, for each transistor, the effect of the other 2 arrangements that have not been employed originally
\item Maintain only the arrangements that result in incorrect performance and thus promote obfuscation
\end{enumerate}

The three-step process described above can be improved if certain configurations of transistors are prioritized. First, it is beneficial to convert transistors with multiple fingers to a single finger wherever possible. This step magnifies performance shifts in transistors due to layout-based effects. Second, we do not need to examine all transistors exhaustively. One can examine transistors by randomly selecting a transistor, choosing an alternative arrangement for it, and assessing the effect on performance. This can be further improved if, as a starting point, one uses a combination of the designer’s experience and circuit symmetry analysis for selecting transistors. 

Finally, the third step can be modified to stop maintaining arrangements that lead to incorrect performance that is too close to the desired performance. We term the keys that create such scenario as \emph{undesirable} keys. In Section \ref{S3}, we consider an OTA as a case study and implement the three-step procedure for locking it.

%These keys activate/select only one (out of three) arrangement for each obfuscated transistor at a time.

%The following process can be used for implementing these steps. In the first step,  In the third step, \textbf{we remove the arrangements that cause nearly correct performance, decrease the number of arrangements that cause correct performance, and maintain the arrangements that result in incorrect performance as much as possible}.

%Note that (in)correct/(un)desirable performance corresponds to (in)correct/(un)desirable keys. In Section \ref{S3}, we consider an OTA as a case study and implement our procedure for locking the circuit.

%% file: Section3.tex
\section{Case Study: Operational Transconductance Amplifier}
\label{S3}

\begin{figure}[t]
%\vspace*{-4mm}
\centerline{\includegraphics[scale=.4]{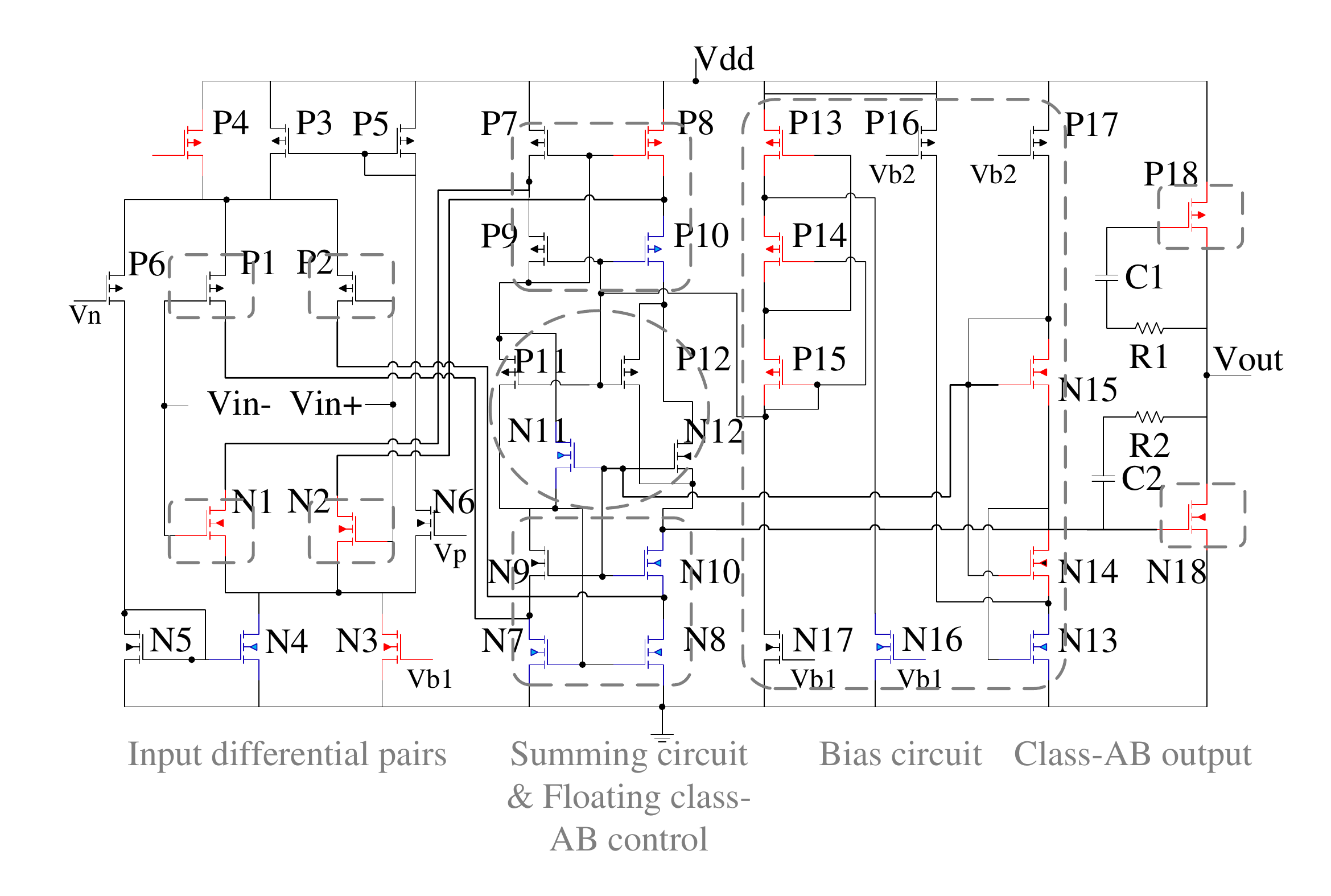}}
\caption{Schematic of OTA circuit. Multiple subcircuits such as input differential pairs (P1, P2, N1, N2), summing circuit (P7-P10, N7-N10), bias circuit (P13-P17, N13-N17), and class-AB output (P18, N18) are used for applying the layout-based effects. Red, blue, and black transistors represent arrangements SOD, SP, and BL, respectively.}
\label{fig: Fig. 6}

\end{figure}

We apply the proposed technique to lock an OTA as shown in Fig. \ref{fig: Fig. 6}. The specs of the OTA for the chosen arrangements are given in Table ~\ref{tab:Table 3}. Note that, in this case study, we only use transistors with a standard voltage threshold. This is, by no means, a limitation of our technique. Then, we examine the effect of unused arrangements on the OTA performance. The number of transistors in the circuit is 36, hence our initial search space is $2^{36 \times 3}$. In practice, we cannot examine all the arrangements for all transistors, but we can explore those with a potential effect on the input differential pairs, summing circuit, floating class-AB control, bias block, and class-AB output (see Fig. \ref{fig: Fig. 6} for details). This type of reasoning is what we previously alluded to when referring to a \emph{designer's expertise} and \emph{symmetry}. 

\begin{figure}[t]
%\vspace*{-3mm}
\centerline{\includegraphics[scale=.4]{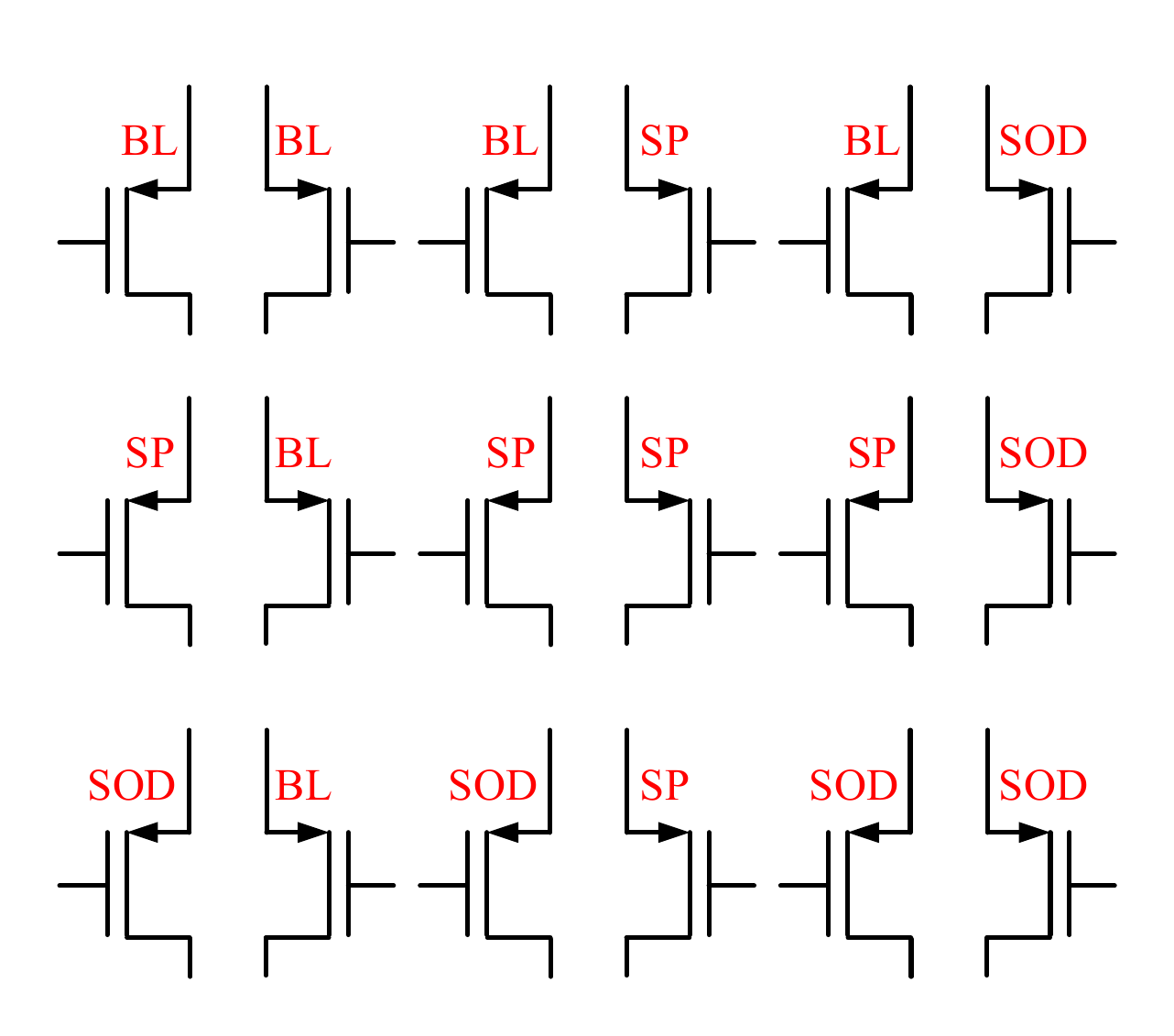}}
%\vspace{-2mm}
\caption{Coupled arrangements. One couple of arrangements is used in the based design as a pair of transistors that can be obfuscated by the other 8 couples of arrangements.}
\label{fig: Fig. 7}
%\vspace{-5mm}
\end{figure}

\begin{table}[t]
%\vspace{-1mm}
	\caption{OTA Specs for utilized arrangements in Fig. \ref{fig: Fig. 6}}
	\begin{center}
		\begin{tabular}{c c c c c}
			\hline
			\hline
			\multicolumn{5}{c}{\sc Specs}    \\
			\hline
			{\sc $gm$} &  {\sc Power*}  & {\sc Gain*} & {\sc Phase} & {3dB \sc Bandwidth}  \\
			\hline
			1.32mS  & 1.1mW   & 73.6dB  & 90deg  & 641KHz \\	
			\hline
			\hline
			
		\end{tabular}
		\vspace{2pt}
		\item \centering *Power is the DC power, and gain is the open-loop gain.
		\label{tab:Table 3}
	\end{center}
	%\vspace{-7mm}
\end{table}

\emph{Simulation results}:  In this work, all simulations are performed by using the Virtuoso Spectre circuit simulator and a commercial 65nm technology. We start with selecting the following $13$ transistors from different subcircuits for obfuscating the circuit: P1, P2, P7-P10, N7-N10, N17, N18, and P18. At this point, the keyspace has $2^{13\times3}$ possible keys. However, not every key is desirable for obfuscating the circuit performance and should be discarded. We shrink the keyspace by applying our three-step procedure as well as controlling a symmetrical pair of transistors, instead of an individual transistor, to balance the layout-dependent effects. The latter technique requires tying together the control bits of the symmetrical pair of transistors. In principle, each symmetrical pair of transistors in the base design can be a couple of arrangements shown in Fig. \ref{fig: Fig. 7}, which in turn can be hidden among other 8 couples of arrangements. In other words, each transistor pair can be obfuscated by at most 8 couples of arrangements. The 13 selected transistors for obfuscation form 6 pairs of transistors plus a singular transistor. We use different number of couples of arrangements for obfuscating these pairs. In total, we added a sum of $28$ random couples of arrangements (i.e., 3, 4, 6, 4, 8, 3) to hide the $6$ pairs of transistors. In addition, we added $1$ single arrangement to obfuscate $1$ transistor. In this experiment, the keylength is $36$ bits (i.e., $28+6+1+1$) made by adding $57$ arrangements (i.e., $28\times2+1$) to the design. At this point, the size of keyspace is $50400$. 

\begin{figure}[t]
%\vspace{-4mm}
\centerline{\includegraphics[width=.65\linewidth]{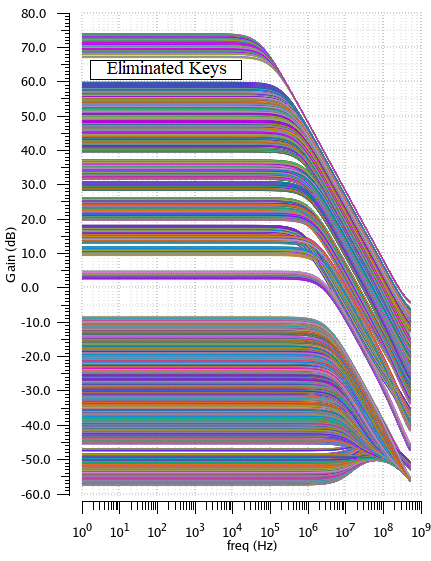}}
 \centering (a)
\centerline{\includegraphics[width=.65\linewidth]{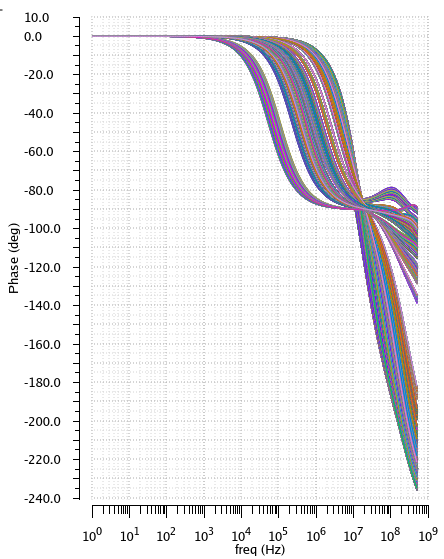}}
\centering (b)
\caption{Layout-based effects on (a) the OTA gain and (b) phase simulated for 50K keys. The gap marked in the graph is the result of purposefully removing nearly correct keys.}
\label{fig: Fig. 8}
%\vspace{-7mm}
\end{figure}

To show the robustness of the achieved obfuscation, we simulated the impact of all different keys on the gain, phase margin, $3dB$ bandwidth (BW), and DC power of the OTA. Fig. \ref{fig: Fig. 8} shows the impact of the keys on the gain. This impact manifests itself as a wide range of degradation in the gain (i.e., up to $130dB$). The desired keys, which generate a gain of $\geq70dB$, may satisfy the design specs. 
In this experiment, the rate of the correct keys, which is adjustable, forms $0.66\%$ of the overall keys. In Fig. \ref{fig: Fig. 8}, there is a gap of $8dB$ between the plots caused by eliminating the nearly correct keys. This is achieved by updating some of the added pairs of arrangements to the circuit. We can further remove the nearly correct keys that produce gain values between $67dB$ and $70dB$. However, these latter keys correspond to less than $0.14\%$ of the overall keys. 

It should be evident that here we establish a trade-off space between keylength and output/behavior “corruption”, which is a dimension also explored in digital logic locking \cite{b24}. Fig. \ref{fig: Fig. 8} also shows the impact of the applied keys on the phase. This impact manifests itself as degradation of up to $50 deg$ in the phase margin. Fig. \ref{fig: Fig. 9} and Fig. \ref{fig: Fig. 10} show the $3dB$ bandwidth and the DC power consumption for the applied keys, respectively. The range of power consumption in the circuit for the correct keys is between $1.143mW$ and $1.188mW$. Interestingly, $1702$ keys cause a power consumption within this range, but only $266$ keys of which are the correct keys. The simulation time for assessing gain and power parameters was nearly $55$ hours long. The area increase as a result of $57$ added arrangements is $158\%$. In addition, power variations of up to $77\%$ compared to that consumed by the initial circuit were noticed. 

As an attempt to protect the correct keys, we apply the following three techniques (and fourth one comes later on): 
\begin{enumerate}
\item Balance the effect of arrangements 
\item Eliminate keys that result in nearly correct performance
\item Remove couples of arrangements with a relatively large impact on performance 

\end{enumerate}

 Let us explain the third technique by an example as the first two techniques have already been discussed. Selecting non-identical couples of arrangements (such as BL-SP or SOD-BL) for N7 and N8 in Fig. \ref{fig: Fig. 6} results in a negative gain, disregarding other arrangements used in the circuit. In other words, there are only three identical couples of arrangements for N7 and N8 to make the gain positive. We eliminate the other six couples of arrangements to remove their alarming effects on the circuit performance. Note that these techniques can result in an uneven number of coupled arrangements for pairs of transistors and raise questions. One attempt to solve this problem is to equalize the number of coupled arrangements for each transistor pair. This solution would shrink the keyspace, but it would regularize the layout and reveal less structural information. We expanded the keyspace by selecting five more transistors: P16-P17 and N13-N15. Now, the locked design has $31$ couples of arrangements (i.e., $28$ minus $6$ removed couples of arrangements from the previous experiment plus $9$ added couples of arrangements in current experiment) to hide $8$ pairs of transistors from the base design. We also kept the single arrangement added to the original design in the previous experiment. Now, the keylength is $41$ bits made by adding $63$ arrangements. We simulated the circuit for the entire keyspace, which has $340200$ keys. Fig. \ref{fig: Fig. 11} shows the impact of these keys on the gain. The target keys form less than $2\%$ of the overall keys. In addition, all gains are positive, and the minimum target gain value is $70dB$. The simulation time for gain, $3dB$ bandwidth and power evaluation was nearly $22$ days. The area is increased by $175\%$ as a result of adding $63$ arrangements. In addition, power variations of up to $73\%$ compared to that consumed in the base circuit were noticed. All simulations were run on a server powered by an Intel (R) Xeon (R) Gold $5122$ CPU with $32$ cores @$3.60GHz$.
 
 The proposed locking scheme can be used for a larger analog circuit than the representative OTA block. In fact, there might not be necessary to apply the locking scheme to all analog blocks. As soon as one of the block is locked, it is very likely that the alteration of the performance of this specific block will affect the performance of the entire circuit. Recall that from the point of view of a single obfuscated transistor in isolation, the overhead is $300\%$. For an analogue circuit, however, only in the absolute worst case would the overhead be $300\%$. We emphasize that not all transistors are obfuscated: some are not satisfactory candidates and some pairs of transistors are jointly obfuscated by fewer combinations of arrangements. State-of-the-art approaches \cite {b12, b13} have displayed smaller overheads while being susceptible to the SMT-based attack \cite{b21}. Our approach establishes a trade-off between overhead and security where we favor higher security. Section~\ref{S4} elaborates the security aspect of the locking scheme.

%% file: Section4.tex
\section{Discussions}
\label{S4}
%%%%%%%%%%%%%%%%%%%%%%%%%%%%%%%%%%%%%%%%%%%%%%%%%%%%%%%%%%%Fig. 8%%%%%%%%%%%%%%%%%%%%%%%%%%%%%%%%
\begin{figure}[t]
%\vspace{-4mm}
\centerline{\includegraphics[width=.95\linewidth]{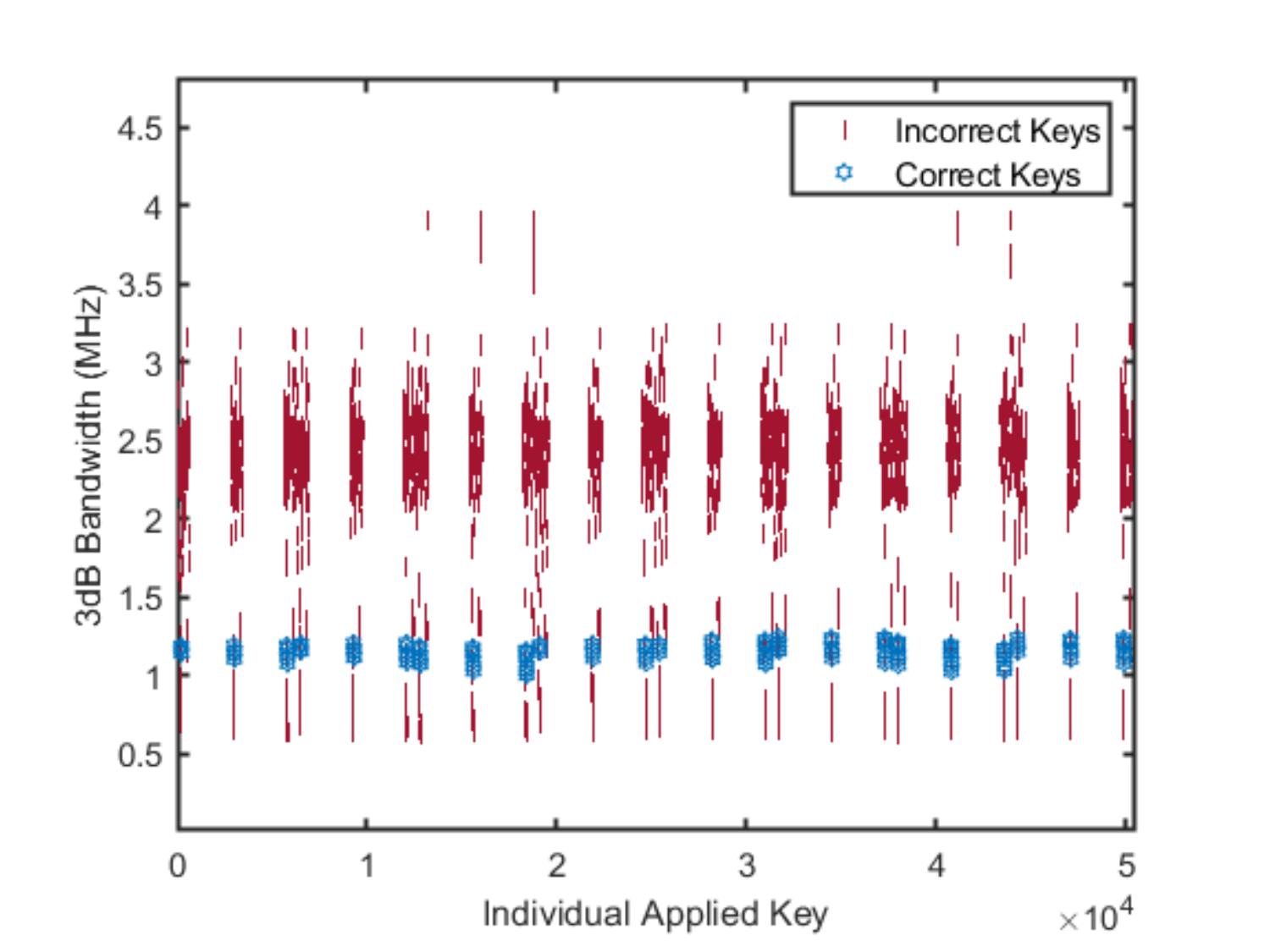}}
%\vspace{-1mm}
\caption{Variation in the 3dB bandwidth of the OTA for the applied keys.}
\label{fig: Fig. 9}
%\vspace{-4mm}
\end{figure}
%%%%%%%%%%%%%%%%%%%%%%%%%%%%%%%%%%%%%%%%%%%%%%%%%%%%%%%%%%%Fig. 9%%%%%%%%%%%%%%%%%%%%%%%%%%%%%%%%
\begin{figure}[t]
%\vspace{-4mm}
\centerline{\includegraphics[width=.95\linewidth]{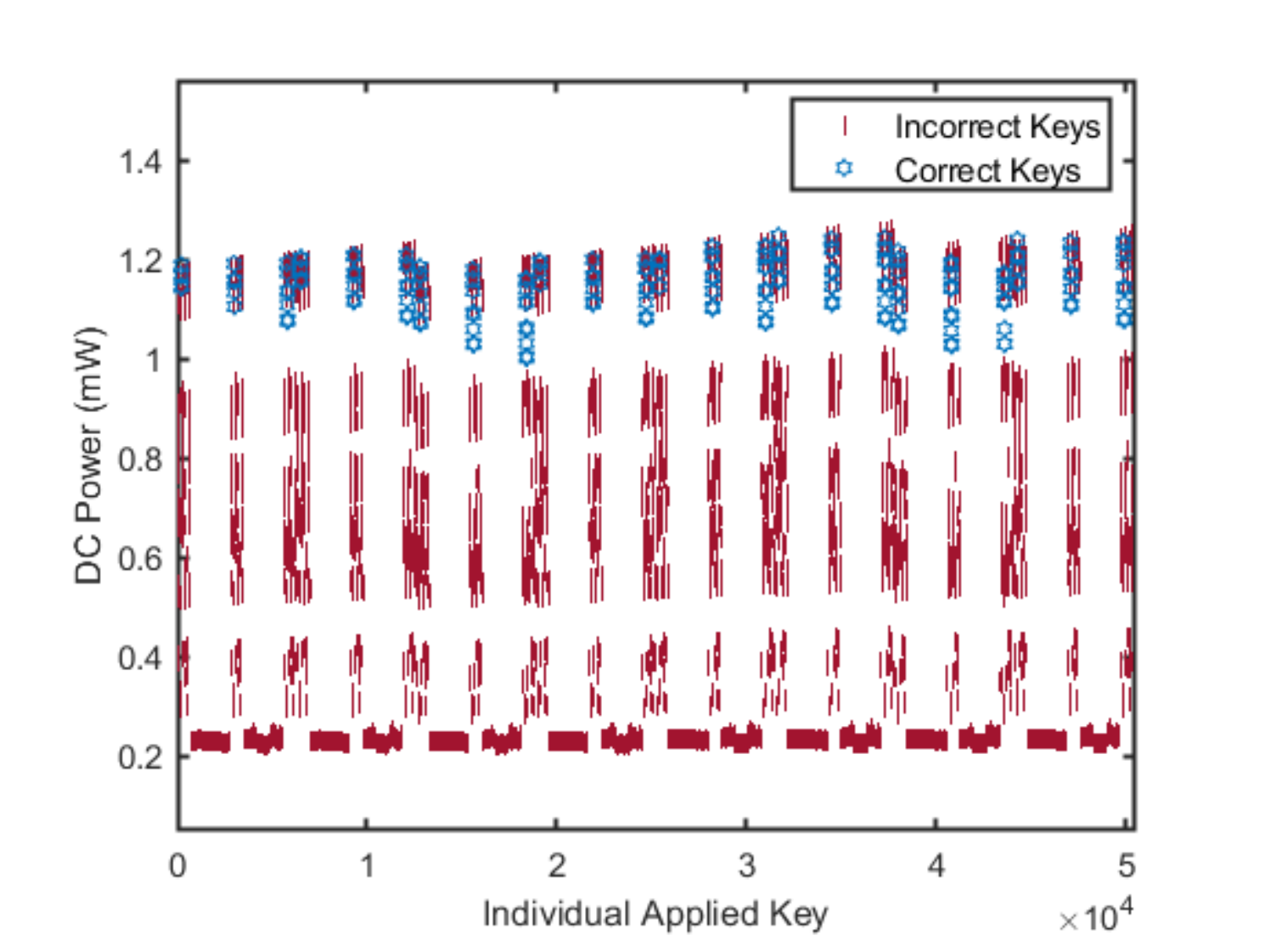}}
%\vspace{-1mm}
\caption{DC power consumption for the applied keys.}
\label{fig: Fig. 10}
%\vspace{-4mm}
\end{figure}
%%%%%%%%%%%%%%%%%%%%%%%%%%%%%%%%%%%%%%%%%%%%%%%%%%%%%%%%%%%Fig. 10%%%%%%%%%%%%%%%%%%%%%%%%%%%%%%%%
\begin{figure}[t]
%\vspace{-4mm}
\centerline{\includegraphics[scale=.55]{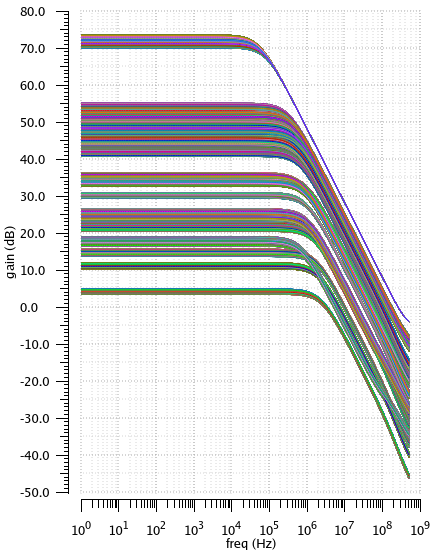}}
%\vspace{-3mm}
\caption{Layout-based effects on the OTA simulated for 300K keys.}
\label{fig: Fig. 11}
%\vspace{-7mm}
\end{figure}

First, we clarify that in our threat model we assume that both foundry and end-user can be untrusted. We assume that the foundry knows every detail about the IP but the correct keys. We assume that a malicious end-user has the required expertise and tools for RE the IP. For example, he/she has access to high-precision optical imaging equipment, circuit simulators, and copies of the functional IP to use as an oracle. However, he/she does not enjoy LDE-level visibility as it is not a current practice in RE efforts. We also assume he/she has no access to a detailed transistor model that takes into account LDEs. In addition, he/she knows to select only one arrangement for each transistor and not more than one. Further, we justify the inefficiency of the brute-force attack, SMT-based attack, and removal attack on the proposed approach. Here we consider the following scenarios for attacking the proposed approach.

\subsection{Untrusted foundry}\label{S4.1}
Everything about the design including LDE-level details is known to the foundry except for the correct keys. As an attempt to further protect the keys, we apply the fourth technique in addition to the three techniques discussed in Section ~\ref{S3}:
\begin{enumerate}
	\setcounter{enumi}{3}
	\item Making the order of the arrangements in the layout design arbitrary.
\end{enumerate}
%\vspace{-2mm}
This technique is to prevent the foundry from trying simple guesses such as ‘all arrangements are BL’ as the order of the key bits due to this technique becomes arbitrary. Given these considerations, we now address the following questions:

%\begin{figure}[h]
%\centerline{\includegraphics[width=0.5\linewidth]{graphs/Fig. 11.pdf}}
%\vspace{-4mm}
%\caption{Layout-based effects on the OTA gain simulated for 300K keys.}
%\label{fig: Fig. 11}
%\vspace{-6mm}
%\end{figure}
\emph{Can a brute force attack compromise the design}? The key sizes considered are within the reach of brute force attacks, specifically for a brute force attack mounted on a real device by observing its performance. Nonetheless, the simulation time in the latter example was $22$ consecutive days for evaluating only $300K$ keys, which are a very small subset of a potentially very large keyspace. For larger circuits and longer keylengths, a brute force attack is not feasible.

\emph{Do partial simulations help to obtain the keys}? Or, in other words, can an adversary decompose the problem into smaller ones and apply a divide and conquer strategy? To answer this question, let us focus on the input differential pairs in the OTA, as an example. An adversary would try different combinations of arrangements for P1, P2, N1, and N2 to find a correct $gm$. He/she would probably find many combinations that result in a correct $gm$. However, most of these combinations will not satisfy other specs of the circuit. The adversary would need to expand the search space to find keys that simultaneously satisfy other specs as well, and this could easily lead to searching the entire keyspace. Another difficulty is that the value of $gm$ depends on the bias circuit, which is also obfuscated, and there might be incorrect bias values that result in the desired value for $gm$. Finding a key that would satisfy, in a deterministic manner, many specs at the same time, does not appear to be feasible.

\emph{Is the SMT-based attack applicable to the proposed approach}? No. The SMT-based attack has been applied to analog ICs with locked bias circuits, where current mirrors or voltage dividers are obfuscated \cite{b12,b13}. The correct key in these circuits is a selection of the mirrored branches, each with a different transistor size, that results in a desirable sum of current. To find the selection, what needs to be done is to write a simple equation, which links the current of the reference branch to the currents of mirrored branches, and assign this task to the SMT solver. The parameters in this equation can be found in the circuit specifications or the PDK (process design kit) documentation. The SMT solver can alone solve this equation without any need for a circuit simulator. This attack has also been applied to a camouflaged analog IP \cite{b15} on the same basis (Table ~\ref{tab:Table 4}). In our approach, however, the layout-based effects are applied to all sub-circuits (e.g., input differential pairs and summing circuit in Fig. \ref{fig: Fig. 6}) and not only to the bias circuit. Therefore, using SMT-based attacks which solve for bias circuits is not sufficient for our approach. Specifically, equations that link the undesirable layout-based effects to circuit performance must be solved by a circuit simulator, and this requirement is \textbf{not scalable}. 

\begin{figure}[t]
\centerline{\includegraphics[width=.96\linewidth]{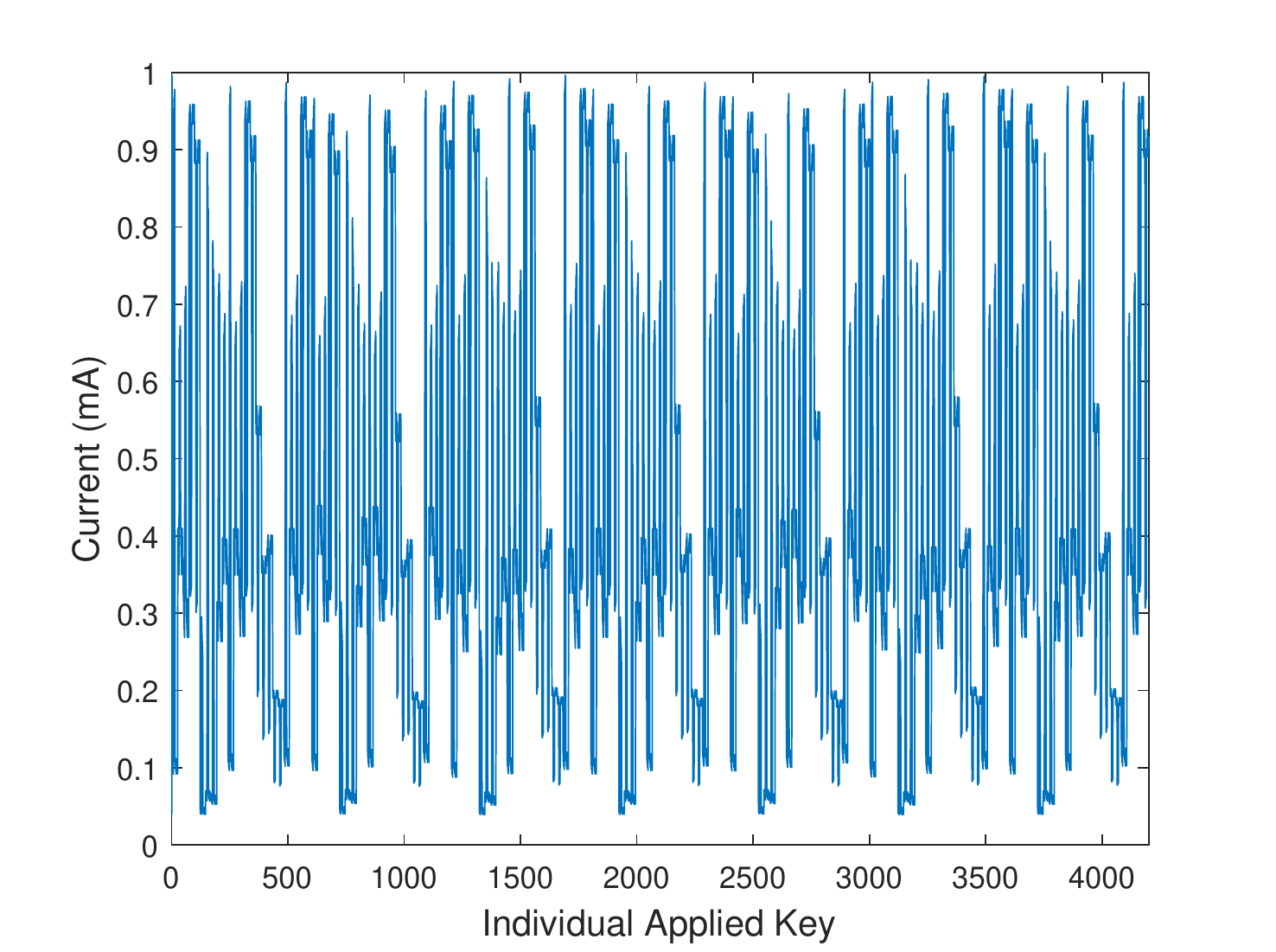}}
\caption{Current variations in an OTA branch for 4K different keys.}
\label{fig: Fig. 12}
\end{figure}

Fig. \ref{fig: Fig. 12} shows a wide range of current variations in one branch of the OTA circuit. The SMT solver should know the desirable range of current in each branch to solve the equations. This cannot be done without extensive simulations. This problem does not exist in the circuits used to apply the SMT-based attack as the currents in those circuit equations are functions of fixed reference currents. Therefore, the existing SMT-based attack cannot be directly applied to our proposed technique. Recently, another attack has been developed for analog biasing locking techniques \cite{b22}. This attack searches for a correct bias instead of the key and is also not applicable to our proposed technique which obfuscates not only the bias circuit but also other parts of the circuit.

\emph{Is the removal attack applicable to the proposed approach}? No. The removal attack is mounted to retrieve the base design by identifying and removing the protection circuitry \cite{b25}. In our locking scheme, the protecting parts cannot be distinguished from the original design. Since our method obfuscate multiple blocks (and not only the biasing block) removing the key-bit transistors would mean redesigning the circuit from the scratch. Specifically, removing the `key-bit transistors' from the OTA would remove $\approx 50\%$ of the original design. In contrast to our method, the state-of-the-art techniques which act on biasing blocks \cite{b11,b12,b13,b16} are vulnerable to a removal attack because the attacker only needs to recover biasing blocks, which typically have a small number of transistors. Although additional steps might be required. In addition, locked AMS designs in \cite{b18,b19} are vulnerable to a removal attack. In fact, the digital lock of the circuits can be removed and the small biasing blocks can be redesigned. Table \ref{tab:Table 4} summarizes the aforementioned discussion and shows the security-overhead trade-off established by our approach. The area overhead in our approach can be reduced to $\sim30\%$ by selecting $2$ (instead of $3$) arrangements per obfuscated transistor at the cost of lowering the security level of the locked circuit.
\begin{table*}[t]
	\caption{Vulnerability of state-of-the-art DfTr methods to SMT-based attack}
	\begin{center}
		\begin{tabular}{c c c c c }
			\hline
			\hline
			 \multirow{2}{*}{\sc DfTr technique} & {\sc Susceptible to} & {\sc Susceptible to} & \multirow{2}{*}{\sc Purely analog} & \multirow{2}{*} {\sc Area overhead (\%)}  \\
			 & {\sc SMT-based attack} & {\sc removal attack}\\
		    \hline
		    \hline
	    	Memristor-based protection  \cite{b11} & Yes & Yes &No & $-$ \\
			\hline
			Parameter biasing obfuscation  \cite{b12} & Yes & Yes & Yes & 6.3  \\
			\hline
			Combinational lock  \cite{b13} & Yes & Yes & Yes & 6.64    \\
			\hline
			Analog Camouflaging  \cite{b15} & Yes & No &Yes & up to 48*   \\
			\hline
			Neural Net. Biasing  \cite{b16} & Yes & Yes & Yes & $-$   \\
			\hline
			AMS lock  \cite{b18} & No  & Yes &No &  0 $\sim$ 171.3**\\
			\hline
			Mix lock  \cite{b19} & No    & Yes &No & 6.7 $\sim$ 24.4** \\
			\hline
			\textbf{This work} & No & No & Yes & 30.6 $\sim$ 175***\\
			\hline
			\hline
		\end{tabular}
		\vspace{2pt}
		\item *This is not a key-based technique, thus the relatively low overhead.
		\item **These values vary depending on the obfuscated circuit and parameters of the locking scheme.
		\item ***Depending on the number of arrangements per transistors selected for obfuscation, which is either 2 or 3, the area overhead varies as shown above. 
		\label{tab:Table 4}
	\end{center}
	%\vspace{-20pt}
\end{table*}

\subsection{Untrusted End-user}\label{S4.2}

Assuming that the netlist is obtained via a RE effort, the IP will not work at the desired performance without knowing the correct key. We assume an adversary can see the metal lines, vias, and even contacts and poly lines of a reverse-engineered circuit. However, the adversary does not enjoy LDE-level visibility. After obtaining the locked netlist, the adversary will see groups of transistors of identical sizes (i.e., our arrangements) since we do not manipulate transistors’ $W$ or $L$. By simulating this obtained netlist with different keys, the adversary will always obtain the same behavior as his model is not detailed enough. And this behavior is not the correct behavior if the circuit was originally designed to exploit LDEs. Even when we assume an adversary possesses an oracle, from which the adversary can confirm that different keys lead to different performance, he has no means to map these shifts back to the design. For readers familiar with the SAT attack \cite{b26}, we highlight that a netlist that is not LDE-aware would prevent an adversary from establishing useful distinguishing input patterns. Therefore, his/her chance of unlocking the circuit is not higher than the malicious foundry, even when possessing an oracle.

%% file: Conclusion.tex
\section{Conclusions}
\label{S5}

This paper shows a novel approach for locking analog ICs. It exploits otherwise undesirable layout-based effects such as WPE and LOD for locking the circuits. This approach is applied to an OTA circuit for a large number of keys to show the robustness of achieved obfuscation. The layout-based effects on gain, phase margin, 3dB bandwidth, and power show the effectiveness of the proposed approach for locking analog ICs. This work demonstrates the potential of the proposed approach for protecting analog circuits against counterfeiting and RE-based attacks.

As a future work, we intend to validate our design in silicon by utilizing a commercial foundry service to achieve a realistic scenario of outsourcing.